\documentclass[a4paper,11pt]{article}

\usepackage{jcappub_nt}
\usepackage[T1]{fontenc} 

\usepackage{bm}
\usepackage{amsmath}

\title{Spatially covariant theories of gravity: disformal transformation, cosmological perturbations and the Einstein frame}

\author[a]{Tomohiro Fujita,}

\author[b,c]{Xian Gao,}

\author[c,d,e]{and Jun'ichi Yokoyama}

\affiliation[a]{Stanford Institute for Theoretical Physics and Department of Physics,\\
                                Stanford University, Stanford, CA 94306, USA}
                        
\affiliation[b]{Department of Physics, Tokyo Institute of Technology,\\
                                2-12-1 Ookayama, Meguro, Tokyo 152-8551, Japan}
                        
\affiliation[c]{Research Center for the Early Universe (RESCEU), Graduate School of Science,\\
                                The University of Tokyo, Tokyo 113-0033, Japan}
                               
\affiliation[d]{Department of Physics, Graduate School of Science,\\
                                The University of Tokyo, Tokyo 113-0033, Japan}
        \affiliation[e]{Kavli Institute for the Physics and Mathematics of the Universe (Kavli IPMU), WPI, UTIAS,
                                The University of Tokyo, Kashiwa 277-8583, Japan}
        
\emailAdd{tomofuji@stanford.edu}

\emailAdd{gao@th.phys.titech.ac.jp}

\emailAdd{yokoyama@resceu.s.u-tokyo.ac.jp}

\abstract{
        We investigate the cosmological background evolution and perturbations in a general class of spatially covariant theories of gravity, which propagates two tensor modes and one scalar mode.
        We show that the structure of the theory is preserved under the disformal transformation.
        We also evaluate the primordial spectra for both the gravitational waves and the curvature perturbation, which are invariant under the disformal transformation.
        Due to the existence of higher spatial derivatives, the quadratic Lagrangian for the tensor modes itself cannot be transformed to the form in the Einstein frame. 
        Nevertheless, there exists a one-parameter family of frames in which the spectrum of the gravitational waves takes the standard form in the Einstein frame.
}

\keywords{}

\arxivnumber{1511.04324}

\begin{document}

\begin{flushright}
	RESCEU-31/15
\end{flushright}
\maketitle

\section{Introduction}

One of the mysteries in modern cosmology is the accelerating expansion of our universe in its primordial epoch \cite{Sato:2015dga}
and today \cite{Riess:1998cb,Perlmutter:1998np}, which stimulates the study of theories containing additional degrees of freedom beyond the simplest model based on general relativity (GR) with a cosmological constant (see \cite{Clifton:2011jh,Joyce:2014kja,Berti:2015itd} for recent reviews).

The most straightforward approach to these additional degrees of freedom is to introduce scalar fields.
Along this direction, a significant progress in the recent years is the re-discovery of the Horndeski theory \cite{Horndeski:1974wa} --- the most general covariant scalar-tensor theory with second derivatives --- as the ``generalized galileon'' \cite{Deffayet:2011gz}.
Lagrangians for a single scalar field with second derivatives were  constructed in a Minkowski background  as the ``galileon'' model \cite{Nicolis:2008in}, which was then generalized to a curved background using the  ``covariantization'' procedure \cite{Deffayet:2009wt,Deffayet:2009mn}.
The generalized galileon  was constructed following the same procedure (see \cite{Deffayet:2013lga} for a review) and
was  shown to be exactly equivalent to the Horndeski theory \cite{Kobayashi:2011nu} (see also \cite{Charmousis:2011bf}).
The Horndeski theory has been applied extensively on the studies of inflation and dark energy, see e.g. \cite{DeFelice:2010pv,Kobayashi:2010cm,Burrage:2010cu,Kobayashi:2011nu,Gao:2011qe,DeFelice:2011uc,Gao:2011mz,Gao:2011vs,DeFelice:2011hq,Tsujikawa:2013ila,deRham:2011by,Heisenberg:2014kea}.

An alternative approach to these additional degrees of freedoms beyond GR, is to construct gravitational theories with less gauge redundancies than those of GR.
The well-studied effective field theory (EFT) of inflation \cite{Creminelli:2006xe,Cheung:2007st,Weinberg:2008hq} (which showed its first appearance in  ghost condensate \cite{ArkaniHamed:2003uy,ArkaniHamed:2003uz}) belongs to this category. 
This approach was further applied to the EFT of dark energy \cite{Creminelli:2008wc,Park:2010cw,Bloomfield:2011np,Battye:2012eu,Gubitosi:2012hu,Bloomfield:2012ff,Mueller:2012kb,Gleyzes:2013ooa,Bloomfield:2013efa,Piazza:2013coa,Piazza:2013pua,Frusciante:2013zop,Gleyzes:2015rua}, 
where the inflaton/dark energy is described by a single scalar degree of freedom, which can also be coupled to several matter fields \cite{Senatore:2010wk,Gwyn:2012mw,Noumi:2012vr,Ballesteros:2013nwa,Gergely:2014rna,Kase:2014yya,Kase:2014cwa,Gleyzes:2015pma} (See \cite{Tsujikawa:2014mba,Kase:2014cwa} for recent reviews).
Another extensively studied example, although initially motivated by a different purpose, is the Ho\v{r}ava gravity \cite{Horava:2009uw} and its healthy extension \cite{Blas:2009qj}.
Interestingly, when fixing the gauge by choosing the scalar field as the
time coordinate ({\em i.e.}, $t = \phi(t,\vec{x})$, which is often dubbed as the ``unitary gauge''), the Horndeski Lagrangian can also be recast in a form similar to that of the EFT of inflation/dark energy \cite{Gleyzes:2013ooa}.
In all of these theories, the Lagrangians involve geometric quantities associated with the foliation of spacelike hypersurfaces only, which respect the spatial diffeomorphism.
Therefore, we may think of such kind of theories as spatially covariant theories of gravity. 

These two approaches have virtually been studied separately.
Nevertheless, recent attempts trying to relate these approaches have brought us new insights into the modification of gravity.
This stems from the observation in \cite{Gleyzes:2014dya} that spatially covariant gravity theories can be used to generate covariant scalar-tensor theories beyond the Horndeski domain, namely theories generally having higher order equations of motion without introducing extra degrees of freedom other than those of the Horndeski theory.
An explicit example was given in \cite{Gleyzes:2014dya}  by minimally deforming the Horndeski Lagrangian in the unitary gauge, of which the healthiness was proven in the unitary gauge \cite{Gleyzes:2014qga,Lin:2014jga} and recently in a covariant manner \cite{Deffayet:2015qwa}.
Such ``beyond Horndeski'' theories have been studied concerning various topics, such as cosmological perturbations \cite{DeFelice:2015isa,Tsujikawa:2015mga}, non-Gaussianities \cite{Fasiello:2014aqa}, Vainshtein mechanism \cite{Kobayashi:2014ida,Saito:2015fza}, etc.

 From the point of view of spatial symmetries, it is natural to explore such ``beyond Horndeski'' theories as general as we can.
This possibility was developed in \cite{Gao:2014soa}, where a wide class of spatially covariant gravity theories was proposed. Through a thorough Hamiltonian analysis, it was shown that such spatially covariant gravity theories propagate at most three degrees of freedom \cite{Gao:2014fra}.
One may thus expect the existence of an even larger class of scalar tensor theories that have higher order equations of motion while still propagating no more than three degrees of freedom.
By construction, the theory in \cite{Gao:2014soa} has virtually included all the previous models.

In fact, scalar-tensor theories beyond the Horndeski domain have been discussed in \cite{Bettoni:2013diz,Zumalacarregui:2013pma} through the so-called ``disformal transformation'' \cite{Bekenstein:1992pj}: $g_{\mu\nu} \rightarrow \Omega^2\, g_{\mu\nu} + \Gamma\, \partial_{\mu}\phi\partial_{\nu}\phi$ where $\Omega$ and $\Gamma$ can be general functions of the scalar field $\phi$ and its derivatives.
Similar to the conformal transformation, the disformal transformation is useful to clarify the mixing between the scalar field $\phi$ and extra matter fields \cite{Zumalacarregui:2010wj,Koivisto:2012za,vandeBruck:2013yxa,vandeBruck:2015ida,vandeBruck:2015rma,vandeBruck:2015tna,Brax:2013nsa,Brax:2014vva,Sakstein:2014isa,Sakstein:2015jca,Bettoni:2015wta,Tsujikawa:2015upa}.%
\footnote{See also \cite{Yuan:2015tta,Huang:2015hja} for recent application of disformal transformation on inducing new geometries.}
It has been shown that the Horndeski theory is \emph{closed} ({\it i.e.} the structure is preserved but with possibly different coefficients) under the disformal transformation if $\Omega$ and $\Gamma$ are both functions of $\phi$ only \cite{Bettoni:2013diz,Zumalacarregui:2013pma} (see also \cite{Arroja:2015wpa}).
While the disformal transformation with $\Gamma=\Gamma(\phi,\partial\phi)$ transforms (subsets of) Horndeski theory to the Lagrangians in \cite{Gleyzes:2014dya,Gleyzes:2014qga}, which themselves are closed under such a transformation.
In systems propagating a single scalar degree of freedom, it has been shown that both the gravitational waves and the curvature perturbations are invariant under the disformal transformation \cite{Minamitsuji:2014waa,Deruelle:2014zza,Tsujikawa:2014uza,Watanabe:2015uqa,Motohashi:2015pra,Domenech:2015hka}.
Moreover, in the case with second order equation of motion, the quadratic action for the tensor modes and thus the corresponding primordial spectrum can be transformed to be those in GR under a suitable disformal transformation \cite{Creminelli:2014wna,DeFelice:2014bma,Tsujikawa:2014uza}.

This paper is devoted to the study of background evolution and generation of perturbations during cosmic inflation of the theory proposed
in \cite{Gao:2014soa}, which we call the XG3 theory referring to eXtended Galileon with 3 space covariance since this theory is more general than the generalized Galileon 
but still their field equations are second order in time in the unitary
gauge.
The 3 also represents the maximum number of second order
derivative operators in each term of the action.  This theory is a
subclass of spatially covariant gravity and is more general than what is
called the GLPV theory \cite{Gleyzes:2014dya}.
We also analyze  transformation laws of the XG3 theory
 under the disformal transformation to show that the theory is closed
 under such a transformation and  that both
 tensor and scalar perturbations are invariant
in the cosmological background.

The rest of the paper is organized as follows.
In the next section we first introduce spatially covariant theories of
gravity and
  examine the transformation of the  XG3 theory 
under the disformal transformation, 
and highlight the relation between the XG3 and GLPV theories. In Sec.\ref{sec:cosmo}
we introduce cosmological background and perturbations in this theory.
In the original Horndeski theory, the equations of motion are second order by construction.
As we shall see in Sec.\ref{sec:linearpert}, generally both the tensor
and scalar modes of the perturbations around a flat
Friedmann-Lema\^{i}tre-Robertson-Walker (FLRW) background possess higher
order spatial derivatives, although the time derivatives are kept up to
the second order, which ensures the absence of any unwanted degree of
freedom.
This fact prevents the possibility of transforming the quadratic action of the gravitational waves itself to be the form in the Einstein frame.
Nevertheless, in Sec.\ref{sec:spec} we will show that we are able to find a frame in which the primordial spectrum of the gravitational wave takes the same the form as in the Einstein frame.
Finally Sec.\ref{sec:concl} is devoted to conclusion.

\section{The formalism}

\subsection{Spatially covariant theories of gravity}
\label{Spatially covariant theories of gravity}

Let us start from the Horndeski theory \cite{Horndeski:1974wa}, which is the most general single-field scalar-tensor theory with second-order equations of motion in four-dimensional spacetime.
The action is reformulated in \cite{Deffayet:2011gz,Kobayashi:2011nu}:
        \begin{equation}
        S^{\mathrm{H}}=\int\mathrm{d}^{4}x\sqrt{-g}\left(\mathcal{L}_{2}^{\text{H}}+\mathcal{L}_{3}^{\text{H}}+\mathcal{L}_{4}^{\text{H}}+\mathcal{L}_{5}^{\text{H}}\right),\label{Horndeski}
        \end{equation}
        with
        \begin{eqnarray}
        \mathcal{L}_{2}^{\text{H}} & = & G_{2}\left(X,\phi\right),\label{Horndeski_L2}\\
        \mathcal{L}_{3}^{\text{H}} & = & G_{3}\left(X,\phi\right)\square\phi,\label{Horndeski_L3}\\
        \mathcal{L}_{4}^{\text{H}} & = & G_{4}\left(X,\phi\right)\,{}^{4}\!R+\frac{\partial G_{4}}{\partial X}\left[\left(\square\phi\right)^{2}-\left(\nabla_{\mu}\nabla_{\nu}\phi\right)^{2}\right],\label{Horndeski_L4}\\
        \mathcal{L}_{5}^{\text{H}} & = & G_{5}\left(X,\phi\right)\,{}^{4}\!G^{\mu\nu}\nabla_{\mu}\nabla_{\nu}\phi-\frac{1}{6}\frac{\partial G_{5}}{\partial X}\left[\left(\square\phi\right)^{3}-3\square\phi\left(\nabla_{\mu}\nabla_{\nu}\phi\right)^{2}+2\left(\nabla_{\mu}\nabla_{\nu}\phi\right)^{3}\right],\label{Horndeski_L5}
        \end{eqnarray}
where $G_{n}$ with $n=2,3,4,5$ are general functions of $\phi$ and
$X\equiv -g^{\mu\nu}\partial_\mu\phi\partial_\nu\phi/2$, ${}^{4}\!R$ and ${}^{4}\!G_{\mu\nu}$ are four-dimensional spacetime Ricci scalar and Einstein tensor, respectively.

The presence of the scalar field $\phi$ naturally foliates the spacetime. Thus the Horndeski theory can be reformulated in terms of the intrinsic and extrinsic geometric quantities associated by the foliation specified by $\phi=\mathrm{const}$.
For our purpose, we assume the scalar field $\phi$ has a timelike
gradient, {\it i.e.},
$g^{\mu\nu}\partial_{\mu}\phi\partial_{\nu}\phi<0$, which specifies the
foliation of codimension-one spacelike hypersurfaces, and express
spacetime metric in terms of the ADM variables
 \cite{Arnowitt:1960es,Arnowitt:1962hi}, 
\begin{equation}
\mathrm{d}s^{2}=-N^{2}\mathrm{d}t^{2}+h_{ij}\left(\mathrm{d}x^{i}+N^{i}\mathrm{d}t\right)\left(\mathrm{d}x^{j}+N^{j}\mathrm{d}t\right).\label{ADM}
\end{equation}
The normal vector to the hypersurfaces is
$n_{\mu}=-\partial_{\mu}\phi/\sqrt{-2X}$.
The components of $\nabla_{\nu} n_{\mu}$ parallel and perpendicular to the hypersurfaces correspond to the extrinsic curvature $K_{\mu\nu} = h^{\rho}_{\mu} \nabla_{\rho} n_{\nu}\equiv \mathrm{D}_{\nu} n_{\mu}$  and the acceleration $a_{\mu} = n^{\nu} \nabla_{\nu} n_{\mu} \equiv \mathrm{D}_{\mu} \ln N$ respectively, where  $h_{\mu\nu} = g_{\mu\nu} + n_{\mu}n_{\nu}$ is the induced metric on the hypersurfaces, $\mathrm{D}_{\mu}$ is the intrinsic covariant derivative compatible with  $h_{\mu\nu}$.
Using the Gauss/Codazzi/Ricci relations as well as
        \begin{equation}
        \nabla_{\mu}\nabla_{\nu}\phi=\frac{1}{N}\left(-n_{\mu}n_{\nu}\pounds_{\bm{n}}\ln N+2n_{(\mu}a_{\nu)}-K_{\mu\nu}\right),
        \end{equation}
where $\pounds_{\bm{n}}$ is the Lie derivative with respect to $n^{\mu}$, 
the Horndeski theory (\ref{Horndeski}) can be equivalently written as \cite{Gleyzes:2013ooa}
        \begin{eqnarray}
        S^{\mathrm{H}} & = & \int\mathrm{d}^{4}x\sqrt{-g}\Big(a_{0}\,K+a_{1}\,{}^{3}\!R\,K+a_{2}\,{}^{3}\!R_{\mu\nu}K^{\mu\nu}+b_{1}\,K^{2}+b_{2}\,K_{\mu\nu}K^{\mu\nu}\nonumber \\
        &  & \hspace{5em}+c_{1}\,K^{3}+c_{2}\,KK_{\mu\nu}K^{\mu\nu}+c_{3}\,K_{\nu}^{\mu}K_{\rho}^{\nu}K_{\mu}^{\rho}+d_{0}+d_{1}\,{}^{3}\!R\nonumber \\
        &  & \hspace{5em}+\nabla_{\mu}\mathcal{B}^{\mu}\Big),\label{SH_dec}
        \end{eqnarray}
where ${}^3\!R_{\mu\nu}$ and ${}^3\!R$ are the intrinsic Ricci tensor and scalar, the 10 coefficients $a_0$ etc are given by
        \begin{eqnarray}
        &  & a_{0}=\frac{\partial F_{3}}{\partial N}-2\frac{1}{N}\frac{\partial G_{4}}{\partial\phi},\qquad\qquad2a_{1}=-a_{2}=\frac{1}{N}F_{5}, \label{coeffH1}\\
        &  & -b_{1}=b_{2}=\frac{\partial\left(NG_{4}\right)}{\partial N}+\frac{1}{2N^{2}}\frac{\partial G_{5}}{\partial\phi},\qquad\qquad c_{1}=-\frac{1}{3}c_{2}=\frac{1}{2}c_{3}=-\frac{1}{6}\frac{\partial G_{5}}{\partial N}, \label{coeffH2}\\
        &  & d_{0}=G_{2}+\frac{1}{N^{2}}\frac{\partial F_{3}}{\partial\phi},\qquad\qquad d_{1}=G_{4}-\frac{1}{2N^{2}}\frac{\partial\left(G_{5}-F_{5}\right)}{\partial\phi}, \label{coeffH3}
        \end{eqnarray}
with $F_3$ and $F_5$ related to $G_3$ and $G_5$ through
        \begin{equation}
                \frac{\partial}{\partial N}\left(\frac{F_{3}}{N}\right)=-\frac{G_{3}}{N^{2}},\qquad\frac{\partial}{\partial N}\left(\frac{F_{5}}{N}\right)=\frac{1}{N}\frac{\partial G_{5}}{\partial N},
        \end{equation}
respectively.
Note in the last line of (\ref{SH_dec}), we kept the boundary term $\nabla_{\mu}\mathcal{B}^{\mu}$ (which was not shown explicitly in \cite{Gleyzes:2013ooa}), where
        \begin{eqnarray}
        \mathcal{B}^{\mu} & = & \left[-\frac{1}{N}F_{3}+2G_{4}K+\frac{1}{2N}F_{5}\left(K^{2}-K_{\rho\sigma}K^{\rho\sigma}\right)\right]n^{\mu}\nonumber \\
        &  & -\left(2G_{4}+\frac{1}{N}F_{5}K\right)a^{\mu}+\frac{1}{N}F_{5}a^{\nu}K_{\nu}^{\mu}-\frac{1}{N}\left(G_{5}-F_{5}\right)\,{}^{4}\!G^{\mu\nu}n_{\nu}.\label{SH_bd}
        \end{eqnarray}

In (\ref{SH_dec}), the 10 coefficients $a_0$ etc are controlled by 4 independent functions $G_1,\cdots,G_4$ through (\ref{coeffH1})-(\ref{coeffH3}). 
It was observed in \cite{Gleyzes:2014dya} that, we may slightly relax the relations in (\ref{coeffH1})-(\ref{coeffH3}) while keeping the relative relations among $\{a_1,a_2\}$, $\{b_1,b_2\}$ and $\{c_1,c_2,c_3\}$, and thus write
        \begin{align}
        & a_{0}=A_{3},\qquad -2a_{1}=a_{2}=B_{5},\qquad  b_{1}=-b_{2}=A_{4}, \label{coeffGLPV1}\\
        & c_{1}=-\frac{1}{3}c_{2}=\frac{1}{2}c_{3}=A_{5},\qquad d_{0}=A_{2},\qquad d_{1}=B_{4}, \label{coeffGLPV2}
        \end{align}
where $A_2$, $A_3$, $A_4$, $A_5$, $B_4$ and $B_5$ are 6 general functions of $\phi$ and $N$.
The resulting theory (dubbed as the ``GLPV theory'' in the literature) lies out of the scope of the original Horndeski theory.
Nevertheless, it was argued in \cite{Gleyzes:2014dya} that the theory still propagates the same number of degrees of freedom as the Horndeski theory.
This striking fact was also proved in
\cite{Lin:2014jga,Gleyzes:2014qga} through the Hamiltonian analysis in a
gauge (usually dubbed as the ``unitary gauge'') where the time 
coordinate is $\phi$ itself, and in a covariant manner \cite{Deffayet:2015qwa}.

In the unitary gauge, the degree of freedom of the scalar field $\phi$ is absorbed into the metric, and the resulting theory looks like a theory for metric variables only, but with reduced symmetries.
In our case with timelike $\partial_{\mu}\phi$, the resulting theory respects only spatial diffeomorphism. Thus we may refer to it as \emph{spatially covariant theory of gravity}.
 From this point of view, there is no reason to restrict the theory by
imposing (\ref{coeffGLPV1})-(\ref{coeffGLPV2}).
Indeed, one may consider a large class of Lagrangians in the unitary gauge, which are general scalar invariants under spatial diffeomorphisms.
This possibility was systematically developed in \cite{Gao:2014soa}, in which the action of the following form
        \begin{equation}
                S = \int \mathrm{d}t \mathrm{d}^3 x\, N \sqrt{h} \, \mathcal{L}(t,N,h_{ij}, R_{ij},K_{ij},\mathrm{D}_{i}). \label{S_gen}
        \end{equation}
was considered, where and in the following, we neglect the superscript ``${}^3$'' for short (and thus $R_{ij}$ stands for the spatial components of ${}^3\!R_{\mu\nu}$ in the unitary gauge, etc).
Through a general Hamiltonian analysis, it was proven in \cite{Gao:2014fra} that the action (\ref{S_gen}) indeed propagates no more than three degrees of freedom, which are the same in the Horndeski theory.
Readers who are familiar with the effective theory of inflation/dark energy or Ho\v{r}ava gravity \cite{Horava:2009uw} may also recognize the similarity among these theories.
As an inverse procedure of gauge fixing, the generally covariant version of (\ref{S_gen}) can be obtained by pushing $t$ to a dynamical scalar field.%
\footnote{Thus $K_{\mu\nu}$, ${}^3\!R_{\mu\nu}$ and $a_{\mu}$ are given by derivatives of the scalar field up to the second order as well as the four dimensional Riemann tensor. See discussions around eq.(6)-(8) in \cite{Gao:2014fra} for details.}
The resulting theory thus become a scalar-tensor theory with a single scalar field ($\phi$-language), which possesses higher order equations of motion, but without introducing extra degrees of freedom.

It should not be surprised that the structure of this class of theories acquires much more freedom than that of the Horndeski theory as the symmetry is reduced.
Instead of considering the general Lagrangian (\ref{S_gen}), 
in this work we restrict ourselves to a subset of (\ref{S_gen}),
dubbed XG3 theory where we impose the following two restrictions: 
1) there are no derivatives higher than the second order in the Lagrangian when going into the ``$\phi$-language''   (and thus we neglect spatial derivatives of $K_{ij}$, $R_{ij}$ and $a_i$), and 
2) the number of second order derivative operators does not exceed three in each term, which enable us to exhaust all possible terms. 
The corresponding Lagrangian can thus be written as \cite{Gao:2014soa}
        \begin{equation}
                S= \int\!\mathrm{d}t \mathrm{d}^3x\, N \sqrt{h}\left( \mathcal{K}_{1}+\mathcal{K}_{2}+\mathcal{K}_{3}+\mathcal{V} \right), \label{Sscg}
        \end{equation}
with
        \begin{eqnarray}
        \mathcal{K}_{1} & = & \left(a_{0}+a_{1}R+a_{3}R^{2}+a_{4}R_{ij}R^{ij}+a_{5}a_{i}a^{i}\right)K\nonumber \\
        &  & +\left[\left(a_{2}+a_{6}R\right)R^{ij}+a_{7}R_{k}^{i}R^{jk}+a_{8}a^{i}a^{j}\right]K_{ij},\label{L_cub_K1_ug}\\
        \mathcal{K}_{2} & = & \left(b_{1}+b_{3}R\right)K^{2}+\left(b_{2}+b_{4}R\right)K_{ij}K^{ij}\nonumber \\
        &  & +\left(b_{5}KK_{ij}+b_{6}K_{ik}K_{j}^{k}\right)R^{ij},\label{L_cub_K2_ug}\\
        \mathcal{K}_{3} & = & c_{1}K^{3}+c_{2}KK_{ij}K^{ij}+c_{3}K_{j}^{i}K_{k}^{j}K_{i}^{k},\label{L_cub_K3_ug}
        \end{eqnarray}
and 
        \begin{eqnarray}
        \mathcal{V} & = & d_{0}+d_{1}R+d_{2}R^{2}+d_{3}R_{ij}R^{ij}+d_{4}a_{i}a^{i}\nonumber \\
        &  & +d_{5}R^{3}+d_{6}RR_{ij}R^{ij}+d_{7}R_{j}^{i}R_{k}^{j}R_{i}^{k}\nonumber \\
        &  & +d_{8}Ra_{i}a^{i}+d_{9}R_{ij}a^{i}a^{j}.\label{L_cub_V_ug}
        \end{eqnarray}
where $a_n,b_n,c_n,d_n$ are general functions of $(t,N)$.
As we shall see, this subclass of Lagrangians has virtually included \emph{all} previous models, while still possessing new interesting extensions.
Before proceeding, note that from the EFT point of view, all operators respecting spatial symmetry should arise, and some of which may introduce unwanted degrees of freedom. We have deliberately dropped such operators in our construction.

\subsection{Disformal transformation} \label{sec:dt}

There have been some interests in investigating scalar-tensor theories from the point of  view of the so-called ``disformal transformation'' \cite{Bekenstein:1992pj}:
\begin{equation}
g_{\mu\nu} \rightarrow \hat{g}_{\mu\nu}=\Omega^{2}\left(\phi\right)g_{\mu\nu}+\Gamma\left(\phi,X\right)\partial_{\mu}\phi\partial_{\nu}\phi. \label{dt}
\end{equation}
When the gauge $\phi=\phi(t)$ is accessible, (\ref{dt}) implies the transformation of ADM variables,
\begin{equation}
\hat{N}=\Phi(t,N) N,\qquad\hat{N}_{i}=\Omega^{2}(t) N_{i},\qquad\hat{h}_{ij}=\Omega^{2}(t) h_{ij}, \label{dtadm}
\end{equation}
where we have denoted
\begin{equation}
\Phi^{2}(t,N)=\Omega^{2}(t)-\frac{1}{N^{2}}\Gamma\left(t,N\right)\left(\partial_{t}\phi\right)^{2} \label{Phi_def}
\end{equation}
for short.
Note that (\ref{dt}) is a spacetime covariant transformation. 
Instead of (\ref{dt})
we may alternatively think of (\ref{dtadm})  as the definition of the disformal transformation.
On each spatial hypersurface with fixed $t$, the disformal
transformation 
(\ref{dtadm}) redefines the lapse function $N$ and  rescales
 the spatial metric $h_{ij}$ and the shift vector $N_i$ simultaneously.

Under the transformation (\ref{dt}), the transformation laws
of the extrinsic and spatial curvatures are given by
\begin{eqnarray}
\hat{K}_{ij} & = & \frac{\Omega^{2}}{\Phi}\left(K_{ij}+\omega\, h_{ij}\right),\qquad\text{with}\qquad\omega\equiv\frac{1}{\bar{N}}\frac{\mathrm{d}\ln\Omega}{\mathrm{d}t}, \label{dt_Kij}\\
\hat{R}_{ij} & = & R_{ij}, \label{dt_Rij}
\end{eqnarray}
and the acceleration transforms as
\begin{equation}
\hat{a}_{i}=\beta a_{i},\qquad\text{with}\qquad \beta\equiv N\frac{\partial\ln\left(\Phi N\right)}{\partial N}. \label{dt_ai}
\end{equation}
Using the above relations, it is straightforward to show the  action (\ref{Sscg}) is \emph{closed} under the disformal transformation. The action in the transformed frame is given by $\hat{S}[\hat{g}_{\mu\nu},\hat{a}_n, \cdots]$ in which all the operators and coefficients in (\ref{Sscg}) are replaced by the hatted ones.
By substituting the above relations, one can rewrite the transformed action in terms of  quantities in the original frame.
Then it is found that only the operators which exist in (\ref{Sscg}) appear
and no new operator emerges. Thus, by rearranging the coefficients as follows, we restore the original action (\ref{Sscg}).
\begin{eqnarray}
\hat{a}_{0} & = & \frac{1}{\Omega^{3}}\left[a_{0}-2\omega(3b_{1}+b_{2})+3\omega^{2}(9c_{1}+3c_{2}+c_{3})\right],\label{dta}\\
\hat{a}_{1} & = & \frac{1}{\Omega}\left[a_{1}-\omega(6b_{3}+2b_{4}+b_{5})\right],\\
\hat{a}_{2} & = & \frac{1}{\Omega}\left[a_{2}-\omega(3b_{5}+2b_{6})\right],\\
\hat{a}_{3} & = & \Omega a_{3},\qquad\hat{a}_{4}=\Omega a_{4},\qquad\hat{a}_{5}=\frac{1}{\beta^{2}\Omega}a_{5},\\
\hat{a}_{6} & = & \Omega a_{6},\qquad\hat{a}_{7}=\Omega a_{7},\qquad\hat{a}_{8}=\frac{1}{\beta^{2}\Omega}a_{8},
\end{eqnarray}
for coefficient functions of terms first order in $K_{ij}$,
\begin{eqnarray}
\hat{b}_{1} & = & \frac{\Phi}{\Omega^{3}}\left[b_{1}-\omega(9c_{1}+2c_{2})\right],\qquad\hat{b}_{2}=\frac{\Phi}{\Omega^{3}}\left[b_{2}-3\omega(c_{2}+c_{3})\right],\\
\hat{b}_{3} & = & \frac{\Phi}{\Omega}b_{3},\qquad\hat{b}_{4}=\frac{\Phi}{\Omega}b_{4},\qquad\hat{b}_{5}=\frac{\Phi}{\Omega}b_{5},\qquad\hat{b}_{6}=\frac{\Phi}{\Omega}b_{6},
\end{eqnarray}
for those second order in $K_{ij}$,
\begin{equation}
\hat{c}_{1}=\frac{\Phi^{2}}{\Omega^{3}}c_{1},\qquad\hat{c}_{2}=\frac{\Phi^{2}}{\Omega^{3}}c_{2},\qquad\hat{c}_{3}=\frac{\Phi^{2}}{\Omega^{3}}c_{3},
\end{equation}
for those third order in extrinsic curvatures, and
\begin{eqnarray}
\hat{d}_{0} & = & \frac{1}{\Phi\Omega^{3}}\left[d_{0}-3\omega a_{0}+3\omega^{2}(3b_{1}+b_{2})-3\omega^{3}(9c_{1}+3c_{2}+c_{3})\right],\\
\hat{d}_{1} & = & \frac{1}{\Phi\Omega}\left[d_{1}-\omega(3a_{1}+a_{2})+\omega^{2}(9b_{3}+3b_{4}+3b_{5}+b_{6})\right],\\
\hat{d}_{2} & = & \frac{\Omega}{\Phi}\left[d_{2}-\omega(3a_{3}+a_{6})\right],\qquad\hat{d}_{3}=\frac{\Omega}{\Phi}\left[d_{3}-\omega(3a_{4}+a_{7})\right],\\
\hat{d}_{4} & = & \frac{1}{\Phi\beta^{2}\Omega}\left[d_{4}-\omega(3a_{5}+a_{8})\right],\qquad\hat{d}_{5}=\frac{\Omega^{3}}{\Phi}d_{5},\\
\hat{d}_{6} & = & \frac{\Omega^{3}}{\Phi}d_{6},\qquad\hat{d}_{7}=\frac{\Omega^{3}}{\Phi}d_{7},\qquad\hat{d}_{8}=\frac{\Omega}{\Phi\beta^{2}}d_{8},\qquad\hat{d}_{9}=\frac{\Omega}{\Phi\beta^{2}}d_{9},\label{dtd}
\end{eqnarray}
for those without any time derivatives,
where in the above $\omega$ and $\beta$ are defined in (\ref{dt_Kij}) and (\ref{dt_ai}), respectively. It should be stressed that no new operator appears, which implies that the action (\ref{Sscg}) is \emph{closed} under the disformal transformation (\ref{dtadm}).

Before proceeding, we would like to make some comments on the form of disformal transformations.
From the point of view of spatial covariance, one may freely consider more general class of transformations that generalizes (\ref{dtadm}):
	\begin{equation}
	\hat{N}=\Phi\,N,\qquad\hat{N}_{i}=\Psi^{2}N_{i},\qquad\hat{h}_{ij}=\Omega^{2}h_{ij}, \label{dtex}
	\end{equation}
where $\Phi$, $\Psi$ and $\Omega$ can be general scalar (under spatial diffeomorphism) functions built of
	\begin{equation}
	t, \; N, \; N_i,\; h_{ij},\; K_{ij},\; R_{ij}, \label{dtexf}
	\end{equation}
as well as their spatial derivatives and even time derivatives.
For instance, the case with $\Psi \equiv \Omega = \Omega(t,N)$ has recently been discussed in \cite{Domenech:2015tca}.
Although it might be interesting to investigate such exotic transformations as (\ref{dtex}), in this paper we restrict ourselves to (\ref{dt}) or (\ref{dtadm}), since our theory is \emph{closed} under it, as we have shown in the above. 

On the other hand, under the transformation (\ref{dtex}) and (\ref{dtexf}), starting from the Horndeski theory one may arrive at  many different classes of theories respecting only spatial symmetries. However, in general such theories involve explicit time derivatives on the lapse, shift or spatial curvature, and thus do not take form as in (\ref{S_gen}). In this work we restrict ourselves to theories in the form of (\ref{S_gen}), of which the XG3 theory is the most general one after imposing the two conditions described above (\ref{Sscg}).

\subsection{Relation between the XG3 theory and the GLPV theory}
\label{Relation between the XG3 theory and the GLPV theory}

Now let us consider the relationship between the XG3 theory
 (\ref{Sscg}) and the GLPV theory in light of the disformal transformation.
As discussed in Sec.~\ref{Spatially covariant theories of gravity},
the latter is a subclass of  the former. The GLPV theory is restored, if (\ref{coeffGLPV1}) and (\ref{coeffGLPV2}) are satisfied, and all of the remaining coefficients  vanish:
\begin{equation}
         a_3,a_4,a_5,a_6,a_7,a_8,b_3,b_4,b_5,b_6,
        d_2,d_3,d_4,d_5,d_6,d_7,d_8,d_9 = 0.
\label{zerocoeffs}
\end{equation}
As we shall show below,  if a theory is included in the spatially covariant theories of gravity but lies out of the scope of the GLPV subclass, 
it cannot be transformed to be a GLPV theory through any non-singular disformal transformation (\ref{dt}).
In other words, once the theory is extended and deviated from the GLPV theory, it cannot be brought back to the GLPV theory by the disformal transformation.
This can be seen as follows.

First, if any coefficients in (\ref{zerocoeffs}) is nonzero, 
the corresponding coefficient in the transformed frame must also be nonvanishing,
and thus the transformed theory is not included in the GLPV subclass. Except for $d_2,d_3$ and $d_4$, the disformal transformations of other coefficients in (\ref{zerocoeffs}) are proportional to themselves. 
If any of them is nonvanishing
in the original frame, they must be nonzero in another disformal frame. 
After setting them to zero, the same argument is applicable to $d_2,d_3$ and $d_4$.
Thus a necessary condition to get a transformed theory in the GLPV subclass is to have all the coefficients in (\ref{zerocoeffs}) being vanishing. In the following discussion, the coefficients in (\ref{zerocoeffs}) are assumed to be zero.

Second, let us consider $a_1$ and $a_2$. The GLPV theory satisfies the
relation $2a_1 = -a_2$. However, $a_1$ and $a_2$ transform in the
exactly same way and the ratio $a_1/a_2$ does not change. Thus if the
relation $2a_1 = -a_2$ is not satisfied in one frame, it cannot be
satisfied in any disformal frame.
For $c_1, c_2$ and $c_3$, one can apply the same argument. Hereafter, we assume that these coefficients satisfy these two relations.

Finally, the remaining parameters are $b_1$ and $b_2$. In the GLPV theory, they satisfy $b_1= -b_2$. With the assumptions made above, their transformations are given by
\begin{equation}
\hat{b}_1 = \frac{\Phi}{\Omega^3} (b_1 - 3\omega c_1),
\qquad
\hat{b}_2 = \frac{\Phi}{\Omega^3} (b_2 + 3\omega c_1).
\end{equation}
Although for a large $\omega$, the transformed theory approaches the GLPV theory, it never reaches the GLPV theory.
In order to have exactly $\hat{b}_1 = -\hat{b}_2$ in the transformed frame, we must have $b_1 = -b_2$ in the original frame.
Therefore, any spatially covariant theory of gravity that does not belong to the GLPV subclass cannot be reduced to the GLPV theory under the disformal transformation (\ref{dt}).

It is interesting to note that since the GLPV theory is \emph{closed}
under the disformal transformation, if a generalized theory could be
reduced to the GLPV theory by the disformal transformation, the GLPV
theory would be the \emph{fixed point} of the transformation.
We have shown, however, that an extended theory cannot 
reach the (would-be) fixed point. Thus the GLPV theory and the rest of 
the XG3 theory in (\ref{Sscg}) are \emph{disconnected} in the space of
theories, in the sense that they are not transformed into each other by the disformal transformation (\ref{dt}).%
\footnote{In contrast, the two pieces of GLPV theory which are beyond Horndeski can be reduced to the form of Horndeski theory by disformal transformations~\cite{Gleyzes:2014qga}. }

\section{Cosmological considerations} \label{sec:cosmo}

In order to apply the XG3 theory to cosmology 
first we separate ADM variables $N$, $N_{i}\equiv h_{ij}N^j$, and
$h_{ij}$  
to the homogeneous part and perturbations as follows.
\begin{eqnarray}
N & = & \bar{N} e^{A},\label{N_A}\\
N_{i} & = & \bar{N} a B_{i},\label{Ni_B}\\
h_{ij} & = & a^{2}\left(e^{\bm{H}}\right)_{ij} \equiv a^2 \left(\delta_{ij}+H_{ij}+\frac{1}{2}H_{i}^{\phantom{i}k}H_{kj}+\cdots\right),\label{hij_exp}
\end{eqnarray}
where $\bar{N}$ and $a$ are functions of $t$.
As usual, we further decompose $B_i$ and $H_{ij}$ into irreducible parts as
\begin{eqnarray}
B_{i} & \equiv & \partial_{i}B+S_{i},\label{Bi_dec}\\
H_{ij} & \equiv & 2\zeta\,\delta_{ij}+\left(\partial_{i}\partial_{j}-\frac{1}{3}\delta_{ij}\partial^{2}\right)E+\partial_{(i}F_{j)}+\gamma_{ij},\label{Hij_dec}
\end{eqnarray}
with $\partial^{2}\equiv\delta^{ij}\partial_{i}\partial_{j}$ and
$\partial^{i}S_{i}=\partial^{i}F_{i}=\partial^{i}\gamma_{ij}=\gamma_{\phantom{i}i}^{i}=0$,
and thus $\zeta\equiv\frac{1}{6}H^{i}_{\phantom{i}i}$ is identified
as the scalar mode and $\gamma_{ij}$ the tensor modes. 
The extrinsic curvature and the acceleration are expressed as:
\begin{eqnarray}
K_{ij} & = & \frac{1}{2N}\left(\dot{h}_{ij}-\mathrm{D}_{i}N_{j}-\mathrm{D}_{j}N_{i}\right),\label{Kij_ADM}\\
a_{i} & = & \partial_{i}\ln N.\label{acce_ADM}
\end{eqnarray}

Now let us consider the change of perturbations under the disformal transformation (\ref{dtadm}). (\ref{hij_exp}) implies that
\begin{equation}
\big(e^{\hat{\bm{H}}}\big)_{ij}=\left(\Omega\frac{a}{\hat{a}}\right)^{2}\left(e^{\bm{H}}\right)_{ij},
\end{equation}
and thus if we \emph{define} the transformation of the scale factor to be
\begin{equation}
\hat{a}\equiv\Omega\,a ,
\label{transformation of scale factor}
\end{equation}
we have $\hat{H}_{ij} = H_{ij}$, and thus
\begin{equation}
\hat{\zeta}\equiv\zeta,\qquad\hat{\gamma}_{ij}\equiv\gamma_{ij}, 
\end{equation}
which implies that the curvature perturbation and the gravitational waves themselves are invariant.
On the other hand, $A$ and $B_i$ are not invariant. Since $N=\bar{N}e^{A}$
and $\hat{N}=\hat{\bar{N}}e^{\hat{A}}$, (\ref{dtadm}) implies $\hat{A}$ is nonlinear in $A$ and is given by
\begin{equation}
\hat{A}=\alpha_{1}A+\frac{1}{2}\alpha_{2}A^{2}+\mathcal{O}\left(A^{3}\right), \label{gdt_A}
\end{equation}
where we have denoted
\begin{equation}
\alpha_{1}\equiv\frac{1}{\bar{\Phi}}\left.\frac{\partial\left(\Phi N\right)}{\partial N}\right|_{N=\bar{N}},\qquad\alpha_{2}\equiv\bar{N}\left.\frac{\partial}{\partial N}\left(\frac{1}{\Phi}\frac{\partial\left(\Phi N\right)}{\partial N}\right)\right|_{N=\bar{N}}, \label{alpha12}
\end{equation}
for short.
We also have
\begin{equation}
\hat{B}_{i}=\Omega^{2}\frac{\bar{N}a}{\hat{\bar{N}}\hat{a}}B_{i}=\frac{\Omega}{\bar{\Phi}}B_{i}.
\end{equation}

\subsection{Background evolution}

By definition, the vanishing of the linear variation of the action
        \begin{equation}
                0=S_{1}=\int\!\mathrm{dt}\mathrm{d}^{3}x\,\bar{N} a^{3}\left(\bar{\mathcal{E}}_{A} A+\bar{\mathcal{E}}_{\zeta}\,3\zeta\right), \label{S1}
        \end{equation}
yields the background equations of motion
        \begin{eqnarray}
        0 & = & \bar{\mathcal{E}}_{A}\equiv d_{0}+d_{0}'+3a_{0}'H-3\left(\lambda_{1}-\lambda_{1}'\right)H^{2}-3\left(2\lambda_{2}-\lambda_{2}'\right)H^{3}, \label{bgeom_A}\\
        0 & = & \bar{\mathcal{E}}_{\zeta}\equiv\Gamma_{1}-\frac{1}{3a^{3}\bar{N}}\frac{\mathrm{d}}{\mathrm{d}t}\left(a^{3}\frac{\partial\Gamma_{1}}{\partial H}\right), \label{bgeom_zeta}
        \end{eqnarray}
with
        \begin{eqnarray}
        \lambda_{1} & \equiv & 3b_{1}+b_{2},\label{lambda_1}\\
        \lambda_{2} & \equiv & 9c_{1}+3c_{2}+c_{3},\label{lambda_2}
        \end{eqnarray}
and
        \begin{equation}
        \Gamma_{1}\equiv d_{0}+3a_{0}H+3\lambda_{1}H^{2}+3\lambda_{2}H^{3},\label{Gamma_1}
        \end{equation}
where the Hubble parameter is defined as
        \begin{equation}
                H\equiv\frac{1}{\bar{N}a}\frac{\mathrm{d}a}{\mathrm{dt}}.
        \end{equation}
Throughout this paper, primes over a function $f=f(t,N)$ denote
        \begin{equation}
                f'\equiv\bar{N}\left.\frac{\partial f}{\partial N}\right|_{N=\bar{N}},\qquad f''\equiv\bar{N}^2\left.\frac{\partial^2 f}{\partial N^2}\right|_{N=\bar{N}},
        \end{equation}  
etc.
Only operators proportional to $a_0$, $d_0$, $b_1$, $b_2$, $c_1$, $c_2$ and $c_3$ contribute to the evolution of background.

Now we consider the transformation of the background equations of motion under the disformal transformation.  In the transformed frame, the first order action takes the same structure as (\ref{S1})
\begin{eqnarray}
\hat{S}_{1} & = & \int\!\mathrm{d}t\mathrm{d}^{3}x\,\hat{\bar{N}}\hat{a}^{3}\left(\mathcal{E}_{\hat{A}}\,\hat{A}+\mathcal{E}_{\hat{\zeta}}\,3\,\hat{\zeta}\right)\nonumber \\
& \supset & \int\!\mathrm{d}t\mathrm{d}^{3}x\,\bar{N}a^{3}\left(\bar{\Phi}\Omega^{3}\mathcal{E}_{\hat{A}}\,\alpha_{1}\,A+\bar{\Phi}\Omega^{3}\mathcal{E}_{\hat{\zeta}}\,3\,\zeta\right), \label{dt_S1}
\end{eqnarray}
thus we have
\begin{eqnarray}
\mathcal{E}_{\hat{A}} & = & \frac{1}{\bar{\Phi}\Omega^{3}\alpha_{1}}\mathcal{E}_{A},\label{dt_EA}\\
\mathcal{E}_{\hat{\zeta}} & = & \frac{1}{\bar{\Phi}\Omega^{3}}\mathcal{E}_{\zeta},\label{dt_Ez}
\end{eqnarray}
where $\alpha_1$ is defined in (\ref{alpha12}).
Since the action is closed under the disformal transformation, (\ref{dt_EA})-(\ref{dt_Ez}) imply that the background equations of motion are invariant (up to irrelevant overall factors) under the disformal transformation.

\subsection{Linear perturbations} \label{sec:linearpert}

Now let us turn to perturbations. The quadratic action for the tensor modes $\gamma_{ij}$ is
\begin{equation}
S_{2}^{\gamma}=\int\!\mathrm{d}t\frac{\mathrm{d}^{3}k}{\left(2\pi\right)^{3}}\bar{N}a^{3}\left(\mathcal{G}_{\gamma}\frac{1}{\bar{N}^{2}}\left(\partial_{t}\gamma_{ij}\right)^{2}-\mathcal{W}_{\gamma}\frac{k^{2}}{a^{2}}\gamma_{ij}^{2}\right),\label{S2g}
\end{equation}
where
\begin{eqnarray}
\mathcal{G}_{\gamma} & = & \frac{1}{4}\left[b_{2}+3\left(c_{2}+c_{3}\right)H\right],\label{Gg}\\
\mathcal{W}_{\gamma} & = & \mathcal{W}_{\gamma}^{(0)}+\mathcal{W}_{\gamma}^{(1)}\frac{k^{2}}{a^{2}}, \label{Wg}
\end{eqnarray}
with
\begin{eqnarray}
\mathcal{W}_{\gamma}^{(0)} & = & \frac{1}{4}\left[d_{1}+\left(3a_{1}+a_{2}\right)H+\left(9b_{3}+3b_{4}+3b_{5}+b_{6}\right)H^{2}\right]\nonumber \\
&  & +\frac{1}{8\bar{N}a}\frac{\mathrm{d}}{\mathrm{d}t}\left\{ a\left[a_{2}+\left(3b_{5}+2b_{6}\right)H\right]\right\} ,\label{Wg0}\\
\mathcal{W}_{\gamma}^{(1)} & = & -\frac{1}{4}\left[d_{3}+\left(3a_{4}+a_{7}\right)H\right].\label{Wg1}
\end{eqnarray}
Throughout this work, repeated lower spatial indices are summed by $\delta_{ij}$.
At this point, note that  the equation of motion for $\gamma_{ij}$
contains spatial derivatives up to the fourth order due to the existence of terms $d_3\,R_{ij}R^{ij}$, $a_4\, R_{ij}R^{ij}K$ and $a_7\, R^{i}_{k}R^{jk}K_{ij}$.%
\footnote{In a recent investigation \cite{Akita:2015dda}, it was shown
that one can stabilize the quadratic curvature gravity at the
level of linear perturbations around a FLRW background by
introducing additional constraints, and get similar Lagrangians with higher spatial derivatives only.}

The quadratic action for the scalar modes takes the following general form
\begin{eqnarray}
S_{2}^{\mathrm{S}} & = & \int\!\mathrm{d}t\frac{\mathrm{d}^{3}k}{\left(2\pi\right)^{3}}\bar{N}a^{3}\bigg[m_{AA}A^{2}+g_{\zeta\zeta}\frac{1}{\bar{N}^{2}}\left(\partial_{t}\zeta\right)^{2}+w_{\zeta\zeta}\frac{k^{2}}{a^{2}}\zeta^{2}+\frac{1}{3}f_{A\zeta}k^{2}A\frac{B}{a}+w_{BB}k^{4}\frac{B^{2}}{a^{2}}\nonumber \\
&  & \hspace{7em}+f_{A\zeta}A\frac{1}{\bar{N}}\partial_{t}\zeta+w_{A\zeta}\frac{k^{2}}{a^{2}}A\,\zeta+\frac{2}{3}g_{\zeta\zeta}k^{2}\frac{B}{a}\frac{1}{\bar{N}}\partial_{t}\zeta+w_{B\zeta}\frac{k^{4}}{a^{2}}\frac{B}{a}\zeta\bigg],\label{S2S}
\end{eqnarray}
where the coefficients $m_{AA}, g_{\zeta\zeta}, 
\cdots, w_{B\zeta}$
can be found in Appendix \ref{sec:Ci}.
At this point, note that generally the Lagrangian we are considering possesses 3 degrees of freedom. Nevertheless, it is possible to choose specific parameters such that at the level of linear perturbations around an isotropic and homogeneous background, only two tensor modes are propagating. This possibility was recently studied in \cite{Yajima:2015xva}, where spatial derivatives of the extrinsic curvature $\mathrm{D}_i K_{jk}$ was also considered. While in this work, we suppose the scalar mode is propagating at the level of linear perturbations.

After solving the constraints for
$A$ and $B$, the final quadratic action for $\zeta$ is given by
\begin{equation}
S_{2}^{\zeta}=\int \mathrm{d}t \frac{\mathrm{d}^{3}k}{(2\pi)^3} \bar{N}a^{3}\left(\mathcal{G}_{\zeta}\frac{1}{\bar{N}^{2}}\left(\partial_{t}\zeta\right)^{2}-\mathcal{W}_{\zeta}\frac{k^{2}}{a^{2}}\zeta^{2}\right),\label{S2zeta}
\end{equation}
where
\begin{eqnarray}
\mathcal{G}_{\zeta} & = & -\frac{(g_{\zeta\zeta}-9w_{BB})\left(f_{A\zeta}^{2}-4g_{\zeta\zeta}m_{AA}\right)}{f_{A\zeta}^{2}-36m_{AA}w_{BB}},\label{Gzeta}\\
\mathcal{W}_{\zeta} & = & \mathcal{W}_{\zeta}^{(0)}+\mathcal{W}_{\zeta}^{(1)}\frac{k^{2}}{a^{2}},\label{Wzeta}
\end{eqnarray}
with
\begin{eqnarray}
\mathcal{W}_{\zeta}^{(0)} & = & -w_{\zeta\zeta}-\frac{1}{2\bar{N}a}\frac{\mathrm{d}}{\mathrm{d}t}\Bigg[\frac{a\left(3f_{A\zeta}^{2}w_{B\zeta}+2f_{A\zeta}w_{A\zeta}(g_{\zeta\zeta}-9w_{BB})-12g_{\zeta\zeta}m_{AA}w_{B\zeta}\right)}{f_{A\zeta}^{2}-36m_{AA}w_{BB}}\Bigg],\label{Wzeta0}\\
\mathcal{W}_{\zeta}^{(1)} & = & -\frac{\left(-3f_{A\zeta}w_{A\zeta}w_{B\zeta}+9m_{AA}w_{B\zeta}^{2}+9w_{A\zeta}^{2}w_{BB}\right)}{f_{A\zeta}^{2}-36m_{AA}w_{BB}}.\label{Wzeta1}
\end{eqnarray}
Similar to the tensor modes, higher spatial derivatives arise in the quadratic action when $\mathcal{W}_{\zeta}^{(1)} \neq 0$.
In particular, from (\ref{mAA}), when $d_4\, a_ia^i$, $a_5\,a_ia^i K$ and $a_8\, a^ia^j K_{ij}$ are present in our model, $m_{AA}$ involves $k^2$ and thus $\mathcal{W}_{\zeta}^{(1)}$ itself depends on $k$. 
Similar cases have been studied in the framework of Ho\v{r}ava gravity \cite{Blas:2009qj,Blas:2010hb}.

We will evaluate the primordial spectra of the tensor and scalar modes in the next section. 
At this point, let us examine whether the quadratic Lagrangians (\ref{S2g}) and (\ref{S2zeta}) are invariant  under the disformal transformation.
Since $\hat{\zeta} = \zeta$ and $\hat{h}_{ij}= h_{ij}$ under the disformal transformation, it is expected that their quadratic actions are also invariant.
This can be verified explicitly by checking the transformation of the corresponding coefficients $\mathcal{G}_{\gamma}$, $\mathcal{W}_{\gamma}$ in (\ref{S2g}), and $\mathcal{G}_{\zeta}$, $\mathcal{W}_{\zeta}$ in (\ref{S2zeta}), as performed in \cite{Tsujikawa:2014uza} in the case of the GLPV theory.
However, here we will show that the invariance of quadratic actions $S^{\gamma}_2$ and $S^{\zeta}_2$ is an inevitable consequence, without explicit calculations.
This is already obvious for the gravitational waves. 
Since both the fully nonlinear action (\ref{Sscg}) and the 
perturbative quantity $h_{ij}$ are invariant, the 
quadratic order action for tensor mode must be 
invariant:
\begin{equation}
\hat{S}_{2}^{\gamma} \equiv  S_{2}^{\gamma},
\end{equation} 
and thus
\begin{equation}
\hat{\mathcal{G}}_{\gamma}=\frac{\bar{\Phi}}{\Omega^{3}}\mathcal{G}_{\gamma},\qquad\hat{\mathcal{W}}_{\gamma}=\frac{1}{\bar{\Phi}\Omega}\mathcal{W}_{\gamma},
\end{equation}
which can be checked explicitly using (\ref{dta})-(\ref{dtd}).

On the other hand, under the disformal transformation, since $A$ and $B_i \equiv \partial_{i} B$ themselves are not invariant, after solving their constraint equations in (\ref{S2S}), it is not manifest if the final quadratic action for the curvature perturbation $S^{\zeta}_{2}$ in (\ref{S2zeta}) is invariant or not.
In particular, the nonlinear transformation of $A$ implies that $\hat{S}_2^{\mathrm{S}} \neq S_2^{\mathrm{S}}$, since $S^{\mathrm{S}}_2$ will contribute to all orders $\hat{S}^{\mathrm{S}}_{n}$ with $n\geq 2$ in the transformed frame.
Nevertheless, following the same procedure as in (\ref{dt_S1}), up to the quadratic order in perturbations, we have the following relation
\begin{eqnarray}
\hat{m}_{\hat{A}\hat{A}} & = & \frac{1}{\bar{\Phi}\Omega^{3}\alpha_{1}^{2}}m_{AA},\qquad\hat{g}_{\hat{\zeta}\hat{\zeta}}=\frac{\bar{\Phi}}{\Omega^{3}}g_{\zeta\zeta},\qquad\hat{w}_{\hat{\zeta}\hat{\zeta}}=\frac{1}{\bar{\Phi}\Omega}w_{\zeta\zeta}, \label{hatco1}\\
\hat{f}_{\hat{A}\hat{\zeta}} & = & \frac{1}{\alpha_{1}\Omega^{3}}f_{A\zeta},\qquad\hat{w}_{\hat{B}\hat{B}}=\frac{\bar{\Phi}}{\Omega^{3}}w_{BB},\qquad\hat{w}_{\hat{A}\hat{\zeta}}=\frac{1}{\alpha_{1}\bar{\Phi}\Omega}w_{A\zeta},\qquad\hat{w}_{\hat{B}\hat{\zeta}}=\frac{1}{\Omega}w_{B\zeta}, \label{hatco2}
\end{eqnarray}
where $\alpha_1$ is defined in (\ref{alpha12}).
Using these results, it is straightforward to verify that
\begin{equation}
\frac{\Omega^{3}}{\bar{\Phi}}\hat{\mathcal{G}}_{\hat{\zeta}}=\mathcal{G}_{\zeta},\qquad\bar{\Phi}\Omega\hat{\mathcal{W}}_{\hat{\zeta}}=\mathcal{W}_{\zeta}, \label{dt_GWzeta}
\end{equation}
where $\hat{\mathcal{G}}_{\hat{\zeta}}$ and $\hat{\mathcal{W}}_{\hat{\zeta}}$ take the same structure with (\ref{Gzeta}) and (\ref{Wzeta}) but with all quantities replaced by the hatted ones in (\ref{hatco1})-(\ref{hatco2}).
It is now manifest from (\ref{dt_GWzeta}) that
\begin{equation}
\hat{S}^{\zeta}_2 \equiv S^{\zeta}_2,
\end{equation}
which implies the quadratic action for the curvature perturbation is invariant under the disformal transformation.

\section{Primordial spectra} \label{sec:spec}

\subsection{Gravitational waves}

Let us calculate the power spectrum of the tensor mode.
By introducing the conformal time $\tau$ through
        \begin{equation}
        \bar{N}\mathrm{d}t=a\,\mathrm{d}\tau, \label{taudef}
        \end{equation}
the quadratic action (\ref{S2g}) can be rewritten as
        \begin{equation}
                S_{2}^{\gamma}=\int\!\mathrm{d}\tau\frac{\mathrm{d}^{3}k}{\left(2\pi\right)^{3}}\frac{1}{2}z^{2}\left[\left(\partial_{\tau}\gamma_{ij}\right)^{2}-\left(1+\frac{k^{2}}{a^{2}M^{2}}\right)c_{\gamma}^{2}k^{2}\gamma_{ij}^{2}\right],
                \label{Sgamma2}
        \end{equation}
where we have defined
        \begin{equation}
        z^{2}=2a^{2}\mathcal{G}_{\gamma}, \qquad c_{\gamma}^{2}=\frac{\mathcal{W}_{\gamma}^{(0)}}{\mathcal{G}_{\gamma}},\qquad M^{2}=\frac{\mathcal{W}_{\gamma}^{(0)}}{\mathcal{W}_{\gamma}^{(1)}}. \label{zcM}
        \end{equation}

In order to canonically quantize the system, we write
        \begin{equation}
        \hat{\gamma}_{ij}\equiv\frac{1}{z}\sum_{s=\pm2}\left[u(\tau,\bm{k})e_{ij}^{(s)}(\hat{\bm{k}})\hat{a}_{s}(\bm{k})+u^{\ast}\left(\tau,-\bm{k}\right)e_{ij}^{(s)\ast}(-\hat{\bm{k}})\hat{a}_{s}^{\dagger}(-\bm{k})\right],
        \end{equation}
where an asterisk denotes complex conjugate, $\hat{a}(\bm{k})$ and $\hat{a}^{\dagger}(\bm{k})$ are the annihilation and creation operators with the commutation relation
        \begin{equation}
        \left[\hat{a}_{s}(\bm{k}),\hat{a}_{s'}^{\dagger}(\bm{k}')\right]=\left(2\pi\right)^{3}\delta_{ss'}\delta^{3}\left(\bm{k}-\bm{k}'\right),
        \end{equation}
$e_{ij}^{(s)}(\hat{\bm{k}})$ is the polarization tensor with the helicity states $s=\pm2$, satisfying
        \begin{equation}
        \sum_{i}e_{ii}^{(s)}(\hat{\bm{k}})=\sum_{i}k^{i}e_{ij}^{(s)}(\hat{\bm{k}})=0, \qquad e_{ij}^{(s)\ast}(\hat{\bm{k}})=e_{ij}^{(-s)}(\hat{\bm{k}})=e_{ij}^{(s)}(-\hat{\bm{k}}).
        \end{equation}
By choosing the normalization
\begin{equation}
\sum_{i,j}e_{ij}^{(s)}(\hat{\bm{k}})e_{ij}^{(s')\ast}(\hat{\bm{k}})=\delta^{ss'},
\end{equation}
$z\hat{\gamma}_{ij}$ is canonically normalized, of which the mode function $u(\tau,\bm{k})$ satisfies the equation of motion
\begin{equation}
        \partial_{\tau}^{2}u(\tau,\bm{k})+\left[\left(1+\frac{k^{2}}{a^{2}M^{2}}\right)c_{\gamma}^{2}k^{2}-\frac{\partial_{\tau}^{2}z}{z}\right]u(\tau,\bm{k})=0. \label{eomu_tau}
\end{equation}
Mode solutions with this kind of modified dispersion relation has been studied in \cite{Martin:2002kt,Ashoorioon:2011eg}.
The two-point function of $\hat{\gamma}_{ij}$ can now be computed as
        \begin{equation}
        \left\langle \hat{\gamma}_{ij}(\bm{k})\hat{\gamma}_{i'j'}(\bm{k}')\right\rangle =\left(2\pi\right)^{3}\delta^{3}\left(\bm{k}+\bm{k}'\right)\mathcal{P}_{ij,i'j'}(\bm{k}),
        \end{equation}
with
\begin{equation}
\mathcal{P}_{ij,i'j'}=\frac{1}{z^{2}}\left|u(\tau,k)\right|^{2}\sum_{s=\pm2}e_{ij}^{(s)}(\hat{\bm{k}})e_{i'j'}^{(s)\ast}(\hat{\bm{k}}).
\end{equation}
The total power spectrum of the gravitational waves is given by
        \begin{equation}
        \mathcal{P}_{\gamma}(k)\equiv\frac{k^{3}}{2\pi^{2}}\mathcal{P}_{ij,ij}=\frac{k^{3}}{\pi^{2}}\frac{1}{z^{2}}\left|u(\tau,k)\right|^{2}. \label{psgen}
        \end{equation}

Now our task is to solve (\ref{eomu_tau}) for the mode function.
Note in the case where all relevant coefficients 
are constant, the background equation of motion (\ref{bgeom_A}) implies
$H$ is constant\footnote{Note this is different from the convention in
\cite{Tsujikawa:2014uza}, where $\bar{N}H$ is chosen to be constant.},
and thus both $c_{\gamma}$ and $M$ are constant, while $z$ is
proportional to $1/\tau$.
Mode function in this case was studied recently in \cite{Yajima:2015xva,Kobayashi:2015gga}.
In this work, we allow the deviation of the background from the exact de Sitter background, which is characterized by the usual slow-roll parameter
        \begin{equation}
                \epsilon_H=-\frac{1}{H^{2}}\frac{\mathrm{d}H}{\bar{N}\mathrm{d}t}. \label{epsilon_def}
        \end{equation}
In this work we assume $\epsilon_H$ to be constant, which implies
        \begin{equation}
                a=a_{\ast}\left(\frac{\tau}{\tau_{\ast}}\right)^{-\frac{1}{1-\epsilon_H}},
        \end{equation}
where $a_{\ast}$ and $\tau_{\ast}$ are constant related by $a_{\ast}=-\frac{1}{\left(1-\epsilon_H\right)H_{\ast}\tau_{\ast}}$ with $H_{\ast}$ is some typical Hubble scale.
Similarly, we allow $c_{\gamma}$, $M$ and $\mathcal{G}_{\gamma}$ to
have some 
time dependence by making the ansatz
\begin{equation}
        c_{\gamma}(\tau)=c_{\gamma\ast}\left(\frac{a}{a_{\ast}}\right)^{\epsilon_{c}},\qquad M(\tau)=M_{\ast}\left(\frac{a}{a_{\ast}}\right)^{\epsilon_{M}},\qquad\mathcal{G}_{\gamma}(\tau)=\mathcal{G}_{\gamma_{\ast}}\left(\frac{a}{a_{\ast}}\right)^{\epsilon_{g}}, \label{epsilon_gamma}
\end{equation} 
where $c_{\gamma\ast}$, $M_{\ast}$ and $\mathcal{G}_{\gamma\ast}$ are
constant, and $\epsilon_c$, $\epsilon_M$ and $\epsilon_{g}$ are constant
numbers. 
These $\epsilon_q~(q=c,M,g)$ parameters are  also given as
        \begin{equation}
        \epsilon_q = \frac{1}{H}\frac{\mathrm{d}\ln q}{\bar{N}\mathrm{d}t}.
        \label{epsilon_gen_def}
        \end{equation}
It should be noted that we do not assume that these slow-roll parameters
are tiny and shall keep the higher order contributions in this section.

It is convenient to introduce a new evolution parameter $x=x(\tau)$ by
        \begin{equation}
        x = -\frac{1-\epsilon_H}{1-\epsilon_H-\epsilon_{c}}c_{\gamma}\tau k.
        \end{equation}
 After some manipulations, the equation of motion for the mode
 function $u$ (\ref{eomu_tau}) is recast in terms of $x$ to
        \begin{equation}
                \frac{\partial^{2}u}{\partial x^{2}}+\frac{1}{x}\left(1+2(\nu-\beta)\right)\frac{\partial u}{\partial x}+\left[1+\alpha^{2}\xi^{2}_\ast x^{2}\left(\frac{x}{x_{\ast}}\right)^{2\epsilon_{\xi}}-\frac{1}{x^{2}}\beta\left(2\nu-\beta\right)\right]u=0, \label{eomu_x}
        \end{equation}
where 
        \begin{equation}
                x_{\ast}=-\frac{1-\epsilon_H}{1-\epsilon_H-\epsilon_{c}} c_{\gamma\ast}\tau_{\ast}k,  \label{xast} 
        \end{equation}
        \begin{eqnarray}
      \alpha & \equiv & 1-\epsilon_H-\epsilon_{c}, \label{alpha}  \\
       \beta & \equiv & \frac{4-2\epsilon_H+\epsilon_{g}}
 {2\left(1-\epsilon_H-\epsilon_{c}\right)},\label{beta}  \\
     \nu & \equiv & \frac{3-\epsilon_H+\epsilon_{g}}{2(1-\epsilon_H-\epsilon_{c})},  \label{nu}
 \end{eqnarray}
and
\begin{equation}
               \epsilon_\xi = \frac{\epsilon_H+\epsilon_{c}+\epsilon_{M}}{1-\epsilon_H-\epsilon_{c}}. \label{pdef}
        \end{equation}
In (\ref{eomu_x}), we have introduced a dimensionless parameter
        \begin{equation}
        \xi(\tau)\equiv\frac{H}{c_{\gamma}M},
        \qquad
        \xi_\ast =\frac{H_{\ast}}{c_{\gamma\ast}M_{\ast}}. \label{xi_def}
        \end{equation}

Unfortunately, (\ref{eomu_x}) does not possess an analytical solution with a nonvanishing $\epsilon_\xi$.
In order to proceed, we assume $\epsilon_\xi \approx 0$.
With this assumption, the solution to (\ref{eomu_x}) is 
        \begin{equation}
                u=x^{\beta}e^{\frac{1}{2}i\alpha\xi_\ast x^{2}}\left[C_{1}U\left(\mu,\nu+1,-i\alpha\xi_\ast x^{2}\right)+C_{2}L_{-\mu}^{\nu}\left(-i\alpha\xi_\ast x^{2}\right)\right],
        \end{equation}
in which $U$ is the confluent hypergeometric function and $L$ is the Laguerre polynomials,
        \begin{equation}
                \mu=\frac{\nu+1}{2}-\frac{i}{4\alpha\xi_\ast},
        \end{equation}
$C_1$ and $C_2$ are two constants to be determined by the initial conditions.
To this end, note when $x\rightarrow +\infty$, the positive frequency mode in the WKB approximation solution to (\ref{eomu_x}) with $\epsilon_\xi=0$ corresponds to
        \begin{equation}
                u\sim e^{-i\int_{x}^{0}\mathrm{d}y\,\alpha\xi_\ast y}= e^{+\frac{1}{2}i\alpha\xi_\ast x^{2}}. \label{wkbuv}
        \end{equation}
On the other hand, since when $x\rightarrow +\infty$,
        \begin{eqnarray}
        U\left(\mu,\nu+1,-i\alpha\xi_\ast x^{2}\right) & \rightarrow & x^{-2\mu}(-i\alpha\xi_\ast)^{-\mu},\\
        L_{-\mu}^{\nu}\left(-i\alpha\xi_\ast x^{2}\right) & \rightarrow & \frac{1}{\Gamma(1-\mu)}x^{-2\mu}(i\alpha\xi_\ast)^{-\mu}\nonumber \\
        &  & +\frac{\Gamma(-\mu+\nu+1)}{\Gamma(1-\mu)\Gamma(\mu)}e^{-i\alpha\xi_\ast x^{2}}x^{2(\mu-\nu-1)}(-i\alpha\xi_\ast)^{\mu-\nu-1},
        \end{eqnarray}
we must have $C_2 = 0$. $C_1$ is fixed by the Wronskian normalization $i=u\frac{\partial u^{\ast}}{\partial\tau}-u^{\ast}\frac{\partial u}{\partial\tau}\equiv\frac{\mathrm{d}x}{\mathrm{d}\tau}\left(u\frac{\partial u^{\ast}}{\partial x}-u^{\ast}\frac{\partial u}{\partial x}\right)$,
which yields
        \begin{equation}
                C_{1}=\frac{1}{\sqrt{2\alpha\xi_\ast c_{\gamma\ast}k}}x_{\ast}^{\frac{\epsilon_{c}}{2\left(1+\epsilon_{c}\right)}}\left(-i\alpha\xi_\ast\right)^{\mu}.
        \end{equation}
The final mode solution is thus given by
        \begin{equation}
                u=\frac{\left(-i\alpha\xi_\ast\right)^{\mu}}{\sqrt{2\alpha\xi_\ast c_{\gamma\ast}k}}\left(x_{\ast}\right)^{\frac{\epsilon_{c}}{2\left(1+\epsilon_{c}\right)}}x^{\beta}e^{\frac{1}{2}i\alpha\xi_\ast x^{2}}U\left(\mu,\nu+1,-i\alpha\xi_\ast x^{2}\right),
                \label{ufin}
        \end{equation}
where $x_{\ast}$ is given by (\ref{xast}).

Since when $x\rightarrow 0$,
        \begin{equation}
                U\left(\mu,\nu+1,-i\alpha\xi_\ast x^{2}\right)\rightarrow x^{-2\nu}\frac{\Gamma(\nu)(-i\alpha\xi_\ast)^{-\nu}}{\Gamma(\mu)},
        \end{equation}
from (\ref{psgen}), the power spectrum on large scales is thus
        \begin{equation}
                \mathcal{P}_{\gamma}=\frac{H_{\ast}^{2}}{\pi^{2}}\frac{\alpha}{4\mathcal{G}_{\gamma\ast}\xi_\ast c_{\gamma\ast}^{3}}\left(x_{\ast}\right)^{n_{\gamma}}\left|\left(-i\alpha\xi_\ast\right)^{\mu-\nu}\frac{\Gamma(\nu)}{\Gamma(\mu)}\right|^{2}, \label{psgamma}
        \end{equation}
with spectral index given by
        \begin{equation}
          n_{\gamma}\equiv\frac{-2\epsilon_H-3\epsilon_c-\epsilon_{g}}{1-\epsilon_H-\epsilon_{c}}. \label{ngamma}
        \end{equation}
It is interesting to note that although our theory contains higher spatial derivatives such that the dispersion relation contains $\sim k^4$ terms, a nearly scale-invariant large scale power spectrum for the gravitational waves can be achieved provided $|n_{\gamma}|\ll 1$.

At this point, one may wonder if the mode solution (\ref{ufin}) and thus the expression for the power spectrum (\ref{psgamma}) are valid only when $\xi \neq 0$, since we have explicitly used a nonvanishing $\xi_{\ast}$ in determining the asymptotic condition (\ref{wkbuv}). 
However, it can be shown explicitly that the mode solution (\ref{ufin}) has a smooth limit of $\xi_{\ast} \rightarrow 0$:
        \begin{equation}
                \lim_{\xi_{\ast}\rightarrow 0} u = e^{i\varphi} u_0, \qquad  u_0\equiv \frac{\sqrt{\pi}}{2\sqrt{c_{\gamma\ast}k}}\left(x_{\ast}\right)^{\frac{\epsilon_{c}}{2\left(1+\epsilon_{c}\right)}}x^{\frac{1}{2\epsilon_{c}+2}}H_{\nu}^{(1)}(x),
        \end{equation}
where $\varphi$ is some $x$-independent phase factor, $u_0$ is the positive frequency mode solution of (\ref{eomu_x}) with $\xi_{\ast} = 0$ and $H^{(1)}_{\nu}$ is the Hankel function of the first kind.
Consequently, $\mathcal{P}_{\gamma}$ in (\ref{psgamma}) can be expanded with respect to $\xi_{\ast}$ as
     \begin{equation}
                \mathcal{P}_{\gamma}= \mathcal{P}_{\gamma 0}\left(1-\mathcal{A}\,\xi_\ast^{2}+\mathcal{O}\left(\xi_\ast^{4}\right)\right),
        \end{equation}
where 
        \begin{equation}
                \mathcal{P}_{\gamma 0}= \frac{H_{\ast}^{2}}{\pi^{2}}\frac{\alpha^{2}}{\mathcal{G}_{\gamma\ast}c_{\gamma\ast}^{3}}\left(\frac{x_{\ast}}{2}\right)^{n_{\gamma}}\frac{\left(\Gamma\left(\nu\right)\right)^{2}}{\pi}
        \label{xizerolimit}
        \end{equation}
is the large-scale power spectrum corresponding to $u_0$, {\it i.e.}, in the absence of $\xi$, and
        \begin{equation}
                \mathcal{A}=\frac{3-\epsilon_H+\epsilon_{g}}{12\left(1-\epsilon_H-\epsilon_{c}\right)}
\left(1+\epsilon_H+2\epsilon_{c}+\epsilon_{g}\right)\left(5-3\epsilon_H-2\epsilon_{c}+\epsilon_{g}\right).
        \end{equation}
It encodes the leading order correction due to the modified dispersion relation.
Due to the assumption of a constant $\xi$ the existence of $k^4$ term  only changes the overall amplitude of the power spectrum, while the spectral index is expected to change if the time dependence of $\xi$ is taken into account.

\subsection{Einstein frame for the gravitational waves}

In this section we show that there exists a one-parameter family of
frames in which the  spectrum of the gravitational waves 
takes the same form as in the GR~\cite{Stewart:1993bc}:
        \begin{align}
        \hat{\mathcal{P}}_\gamma^{\rm GR}(k) &= \frac{2
         \hat{H}^2(k_\ast)}
{\pi^2 M_{\rm Pl}^2}\left(\frac{k}{k_\ast}\right)^{\hat{n}^{\rm GR}_\gamma} [1+\mathcal{E}^{\rm GR}(\hat{\epsilon}_H)],\label{standard_exp}\\
        \hat{n}^{\rm GR}_\gamma&=-\frac{2\hat{\epsilon}_H}{1-\hat{\epsilon}_H},\\
        \mathcal{E}^{\rm GR}(\hat{\epsilon}_H)&=\pi^{-1}2^{\frac{2}{1-\hat{\epsilon}_H}}(1-\hat{\epsilon}_H)^2
        \Gamma^2\left(\frac{3-\hat{\epsilon}_H}{2(1-\hat{\epsilon}_H)}\right)-1,
        \end{align}
under the assumption that the slow-roll parameters are constant and small.
Here $M_{\rm Pl}$ is the reduced Planck mass and $k_\ast$ is an
arbitrary pivot scale where we take $x_\ast=1$.
$\hat{n}^{\rm GR}_\gamma$ and $\mathcal{E}^{\rm GR}$ are the tensor spectral index and the slow-roll correction to the amplitude, respectively.%
\footnote{One can derive them by setting $c_{\gamma*}=1$, $\mathcal{G}_{\gamma*}=M_{\rm Pl}^2/8$, $\epsilon_c=0$, $\epsilon_g=0$ and $k_*=-1/\tau_*$ in (\ref{xizerolimit}).}
Although the sound speed $\hat{c}_\gamma$ and the overall factor $\hat{\mathcal{G}}_\gamma$ are not necessarily the same as the GR case and unfamiliar terms (e.g. the forth spatial derivative term) can exist in the tensor quadratic action in those frames, the power spectrum completely coincide with the one in the GR.
It should be stressed that the argument in this subsection is not restricted to the XG3 theory but applicable to any theory in which the tensor perturbation is invariant under the disformal transformation.

\subsubsection{General argument}
\label{General argument}

Let us consider a general tensor power spectrum in the original frame:
        \begin{equation}
        \mathcal{P}_\gamma^{\rm gen}=\mathcal{P}_{\gamma(0)}^{\mathrm{gen}}
        \left(\frac{k}{k_\ast}\right)^{n_\gamma^{\rm gen}}
        \left( 1 +\mathcal{E}^{\rm gen}\right),
        \label{gen_P}
        \end{equation}
where $\mathcal{P}_{\gamma(0)}^{\mathrm{gen}}$ is the power spectrum at the 0-th order of the slow-roll expansion, $n_\gamma^{\rm gen}$ is the spectral index which is at least the first order in the slow-roll parameters, and  $\mathcal{E}^{\rm gen}$ is the slow-roll correction to the amplitude which includes the first order and higher terms of slow-roll parameters. 
Note that the running of the tensor spectral index vanishes due to the assumption that the slow-roll parameters are constant.

We set the following ansatz to parameterize the disformal transformation
consistently with our assumption;
\begin{equation}
\Omega(t)=\Omega_{\ast}\left(\frac{a}{a_{\ast}}\right)^{\epsilon_{\Omega}},
\qquad
\bar{\Phi}(t)=\bar{\Phi}_{\ast}\left(\frac{a}{a_{\ast}}\right)^{\epsilon_{\bar{\Phi}}},
\qquad
\epsilon_{\Omega},\,\epsilon_{\bar{\Phi}}=\mathrm{const}.
\end{equation}
In what follows, we determine these parameters $\bar{\Phi}_\ast$, $\epsilon_\Omega$ and $\epsilon_{\bar{\Phi}}$ such that (\ref{gen_P}) is mapped to (\ref{standard_exp}) in new frames, while $\Omega_\ast$ remains undetermined.
To this end, we rewrite (\ref{standard_exp}) in terms of quantities in the original frame,
\begin{align}
\hat{\mathcal{P}}_\gamma^{\rm GR} &=\frac{2
         H^2_\ast}
{\pi^2 M_{\rm Pl}^2 \bar{\Phi}_\ast^2}\left(\frac{k}{k_\ast}\right)^{\hat{n}^{\rm GR}_\gamma}  
\Big(1+\epsilon_\Omega\Big)^2\Big[1+\mathcal{E}^{\rm GR}(\epsilon_H+\epsilon_{\bar{\Phi}})\Big], \label{PGR}
\\
\hat{n}_\gamma^{\rm GR}&=-2\frac{\epsilon_H+\epsilon_{\bar{\Phi}}}{1-\left(\epsilon_H+\epsilon_{\bar{\Phi}}\right)},\label{PGRng}
\end{align}
where we have used $\hat{H}_*=(1+\epsilon_\Omega)H_*/\bar{\Phi}_*$ and $\hat{\epsilon}_H=\epsilon_H+\epsilon_{\bar{\Phi}}$.
Using the invariance of the power spectrum $\mathcal{P}_\gamma^{\rm gen}=\hat{\mathcal{P}}_\gamma^{\rm gen}$, the equality between the general power spectrum in a new frame $\hat{\mathcal{P}}_\gamma^{\rm gen}$ in (\ref{gen_P}) and the GR expression $\hat{\mathcal{P}}_\gamma^{\rm GR}$ (\ref{PGR})-(\ref{PGRng}) are decomposed into the following three pieces. \\
(i) The 0-th order amplitude: 
\begin{equation}
\mathcal{P}_{\gamma(0)}^{\mathrm{gen}}=\frac{2
         H^2_\ast}
{\pi^2 M_{\rm Pl}^2 \bar{\Phi}_\ast^2}
\quad\Longrightarrow\quad
\bar{\Phi}_\ast = \frac{\sqrt{2}H_\ast}{\pi M_{\rm Pl}} \left(\mathcal{P}_{\gamma(0)}^{\rm gen}\right)^{-\frac{1}{2}}.
\label{DT0amp}
\end{equation}
(ii) The tensor spectral index:
\begin{equation}
n_\gamma^{\rm gen}=-\frac{2\left(\epsilon_H+\epsilon_{\bar{\Phi}}\right)}{1-\epsilon_H-\epsilon_{\bar{\Phi}}}
\quad\Longrightarrow\quad
\epsilon_{\bar{\Phi}}=-\frac{2\epsilon_H+(1-\epsilon_H)n_\gamma^{\rm gen}}{2-n_\gamma^{\rm gen}}.
\label{DTspeind}
\end{equation}
(iii) The slow-roll correction to the amplitude:
\begin{equation}
1+\mathcal{E}^{\rm gen} =\Big(1+\epsilon_\Omega\Big)^2\Big(1+\mathcal{E}^{\rm GR}\Big)
\quad\Longrightarrow\quad
\epsilon_\Omega =\sqrt{\frac{1+\mathcal{E}^{\rm gen}}{1+\mathcal{E}^{\rm GR}}}-1.
\label{DTsrcrr}
\end{equation}
Note that $\Omega_*$ remains undetermined and it stays as a free parameter.
Therefore we obtain the one-parameter family of frames in which the general tensor power spectrum in the original frame is given by the expression of the GR.

For instance, in the case of the XG3 theory with the vanishing $\epsilon_\xi$, namely (\ref{psgamma}),
the 0-th order amplitude, the tensor spectral index and the slow-roll correction to the amplitude are given by
\begin{align}
\mathcal{P}_{\gamma(0)}^{\rm XG3} &= \frac{H_*^2 \exp[-\pi/4\xi_\ast]}{16\pi \mathcal{G}_{\gamma\ast}c_{\gamma\ast}^3\xi_\ast^{3/2}|\Gamma(5/4+i/4\xi_*)|^2},\notag\\
n_\gamma^{\rm XG3} &= \frac{-2\epsilon_H-3\epsilon_c-\epsilon_{g}}{1-\epsilon_H-\epsilon_{c}},\\
1+\mathcal{E}^{\rm XG3} &= \frac{4\alpha}{\pi}\xi_*^{1/2}e^{-\frac{\pi}{4\xi_*}}\left|(-i\alpha\xi_*)^{\mu-\nu}\frac{\Gamma(\nu)}{\Gamma(\mu)}\Gamma\left(\frac{5}{4}+\frac{i}{4\xi_*}\right)\right|^2.
\notag
\end{align}
Substituting these into (\ref{DT0amp})-(\ref{DTsrcrr}), one obtains the one-parameter family of disformal transformation which maps (\ref{psgamma}) to (\ref{standard_exp}).

\subsubsection{Einstein-frame and transformation}

Here, we would like to make a few comments on the Einstein frame. The Einstein frame normally refers to the frame in which the quadratic action for the tensor mode $S_2^\gamma$ is given by the standard form of the GR. However, from the above analysis, we found that  although there always exists frames in which the tensor power spectrum $\mathcal{P}_\gamma$
is given by the expression of GR (\ref{standard_exp}) under the assumption of the constant slow-roll parameters, the quadratic action $S_2^\gamma$ does not necessarily coincide with that of the GR in these frames.
For instance, the fourth derivative term in (\ref{Sgamma2}) is the explicit deviation from the GR, which cannot be removed by the disformal transformation.\  But yet, the power spectrum (\ref{psgamma}) in the XG3 theory can be mapped into (\ref{standard_exp}). 
Note that the dynamics of the tensor modes is clearly different from that of the GR in those frames. Only the observable  $\mathcal{P}_\gamma$ is the same as the GR case.

At this point, it is worthwhile to consider the case of the GLPV theory.
The quadratic action of the tensor mode in the GLPV theory is
simply (\ref{Sgamma2}) without the fourth derivative term. 
Following the procedure discussed in this subsection, one can determine the one-parameter family of frames, in which the tensor power spectrum
is given by the standard expression (\ref{standard_exp}).
Moreover, in the case of the GLPV theory, one can fix $\Omega_*^{\rm EF}$ for the parameter $\Omega_*$ such that the quadratic action for the tensor modes coincides with that in GR, namely the Einstein frame~\cite{Tsujikawa:2014uza}.
Conversely, for any $\Omega_*$ which is \emph{not} that specific value $\Omega_*^{\rm EF}$, the rest of the one-parameter family of frames does not include the Einstein-frame, while the tensor power spectrum still agrees with that in the GR.

Furthermore, it is interesting to note that even in the presence of the conformal transformation $g_{\mu\nu} \rightarrow \hat{g}_{\mu\nu}=\Omega^{2}\left(\phi\right)g_{\mu\nu}$ only,
the 0-th order amplitude and the spectral index of $\mathcal{P}_\gamma$ can be freely tuned and reduced to the form in GR, while the slow-roll correction to the amplitude does not agree with $\mathcal{E}^{\rm GR}$ in general. It can be seen easily by taking the limits $\Gamma(t,N)\to 0$
and $\Phi(t,N)\to \Omega(t)$ in (\ref{DT0amp})-(\ref{DTsrcrr}), since in this case (\ref{DTspeind}) and (\ref{DTsrcrr}) cannot be satisfied simultaneously. 
It implies that the disformal transformation can map a general tensor power spectrum to the GR, because it contains the two free functions $\Omega(t)$ and $\Gamma(t,N)$, which appear in the transformed tensor power spectrum in such a way that  the slow-roll correction to the amplitude can be adjusted as well.

\subsection{Curvature perturbation}

For completeness, in the following we also evaluate the power spectrum for the curvature perturbation $\zeta$.
As we have discussed, generally $\mathcal{W}_{\zeta}^{(1)}$ may have complex functional dependency on $k$.
In the following, we assume the $k$-dependency in $\mathcal{W}_{\zeta}^{(1)}$ can be neglected.
By writing
        \begin{equation}
                \hat{\zeta}(\tau,\bm{k})=\frac{1}{z_{\zeta}}\left(v(\tau,k)\hat{b}(\bm{k})+v^{\ast}(\tau,k)\hat{b}^{\dagger}(-\bm{k})\right),
        \end{equation} 
where
        \begin{equation}
                z_{\zeta}^{2}=2a^{2}\mathcal{G}_{\zeta},
        \end{equation}
$\hat{b}$ and $\hat{b}^{\dagger}$ satisfy the commutation relation $\left[\hat{b}(\bm{k}),\hat{b}^{\dagger}(\bm{k}')\right]=(2\pi)^{3}\delta^{3}(\bm{k}-\bm{k}')$, 
the equation of motion for the mode function $v(\tau,k)$ is given by 
        \begin{equation}
                \partial_{\tau}^{2}v(\tau,k)+\left[\left(1+\frac{k^{2}}{a^{2}M_{\zeta}^{2}}\right)c_{\zeta}^{2}k^{2}-\frac{\partial_{\tau}^{2}z_{\zeta}}{z_{\zeta}}\right]v(\tau,\bm{k})=0,
        \end{equation}
where $\tau$ is the conformal time defined in (\ref{taudef}), and
        \begin{equation}
                c_{\zeta}^{2}=\frac{\mathcal{W}_{\zeta}^{(0)}}{\mathcal{G}_{\zeta}},\qquad M_{\zeta}^{2}=\frac{\mathcal{W}_{\zeta}^{(0)}}{\mathcal{W}_{\zeta}^{(1)}}.
        \end{equation}
Solving $v$ is completely parallel to that of the tensor modes.
By making the similar ansatz as in (\ref{epsilon_gamma}) for the parameters:
        \begin{equation}
c_{\zeta}=c_{\zeta\ast}\left(\frac{a}{a_{\ast}}\right)^{\tilde{\epsilon}_{c}},
\qquad M_{\zeta}=M_{\zeta\ast}
\left(\frac{a}{a_{\ast}}\right)^{\tilde{\epsilon}_{M}},
\qquad\mathcal{G}_{\zeta}=\mathcal{G}_{\zeta_{\ast}}
\left(\frac{a}{a_{\ast}}\right)^{\tilde{\epsilon}_{g}},
        \end{equation}
and assuming $c_{\zeta\ast}$, $M_{\zeta\ast}$, $\mathcal{G}_{\zeta_{\ast}}$,
$\tilde{\epsilon}_{c}$, $\tilde{\epsilon}_{M}$ and
$\tilde{\epsilon}_{g}$ 
are constant, after some manipulation, 
the large scale power spectrum of the curvature perturbation reads
        \begin{equation}
                \mathcal{P}_{\zeta}=\frac{H_{\ast}^{2}}{8\pi^{2}}\frac{\tilde{\alpha}}{\mathcal{G}_{\zeta\ast}\tilde{\xi}c_{\zeta\ast}^{3}}\left(y_{\ast}\right)^{n_{\zeta}-1}\left|\left(-i\tilde{\alpha}\tilde{\xi}\right)^{\tilde{\mu}-\tilde{\nu}}\frac{\Gamma(\tilde{\nu})}{\Gamma(\tilde{\mu})}\right|^{2},
        \end{equation}
where
        \begin{equation}
y_{\ast}=-\frac{1-\epsilon_H}{1-\epsilon_H-\tilde{\epsilon}_{c}}
\tau_{\ast}c_{\zeta\ast}k,\qquad
\tilde{\xi}=\frac{H_{\ast}}{c_{\zeta\ast}M_{\zeta\ast}},
        \end{equation}
and
        \begin{eqnarray}
        \tilde{\alpha} & \equiv & 1-\epsilon_H-\tilde{\epsilon}_{c},
\label{alpha_t}\\
        \tilde{\nu} & \equiv & \frac{3-\epsilon_H+\tilde{\epsilon}_{g}}
{2(1-\epsilon_H-\tilde{\epsilon}_{c})},\label{nu_t}\\
        \tilde{\mu} & = & \frac{\tilde{\nu}+1}{2}-\frac{i}{4\tilde{\alpha}\,\tilde{\xi}}.\label{mu_t}
        \end{eqnarray}
The spectral index is given by
        \begin{equation}
n_{\zeta}-1 =
 -\frac{2\epsilon_H+3\tilde{\epsilon}_{c}+\tilde{\epsilon}_{g}}
{1-\epsilon_H-\tilde{\epsilon}_{c}}.
        \end{equation}

\section{Conclusion} \label{sec:concl}

In the present paper we have studied some properties of spatially
covariant gravity with a scalar field $\phi$ in cosmological contexts.
While the most general theory of scalar-tensor theory with full
covariance in four spacetime dimension, whose field equations are
second-order differential equations, is known as the generalized
Galileon \cite{Deffayet:2011gz,Kobayashi:2011nu} or the Horndeski theory
 \cite{Horndeski:1974wa}, 
our theory is formulated to
preserve covariance only in a spatial hypersurface which is conveniently
defined by a $\phi=$constant surface \cite{Gao:2014soa}.  Here $\phi$ plays the role of 
time and second-order covariant derivative of $\phi$ can be expressed in
terms of the extrinsic curvature $K_{ij}$ of the hypersurface.
Then the theory can be expressed by a sum of combinations of the geometrical quantities,
and the coefficient of each term is treated as functions of $t$ and the lapse $N$
instead of $\phi$ and $X$.

Specifically as in the case of the generalized Galileon, the theory we
used contains no terms with third or higher order derivatives in $\phi$
and the number of second-order derivatives is limited to three in each
term.  That is why we call it the eXtended Galileon with 3 space
covariance and up to 3 second-order derivatives, in short, the XG3
theory \cite{Gao:2014soa}.
Although it is similar to the GLPV theory \cite{Gleyzes:2014dya} in the sense that it has
spatial covariance only but still the number of propagating modes is
three, the XG3 theory is more general than the GLPV theory.  In
particular, unlike the GLPV theory, cosmological perturbations in the
XG3 theory contain fourth-order spatial derivative terms in the action. 

We have studied how the theory responds to the disformal transformation
\cite{Bekenstein:1992pj}, to  show that both the XG3 and GLPV
theories are  closed under generic disformal transformation, which means
that the GLPV theory constitutes an isolated subclass of the XG3 theory
with respect to this transformation.  

We have also shown the actions for the linear tensor and curvature
perturbations are also invariant with respect to the disformal
transformation.
We have particularly focused on the tensor perturbations generated
during cosmic inflation, which is less dependent on inflation models
than curvature perturbations,  to show that one can find a disformal
transformation which maps both the amplitude and the spectral index of
tensor power spectrum to the form predicted by the general relativity.

It is remarkable that theories containing higher order spatial
derivative terms in the action can realize the tensor power spectrum
in full agreement with the prediction of the Einstein gravity, and
so the utmost care must be devoted in interpreting the observational
data of tensor perturbations.

Last but not least, as discussed in the end of Sec.\ref{sec:dt}, it is interesting to explore more general transformations such as (\ref{dtex}) and (\ref{dtexf}), and investigate their observational signatures, which may help us to distinguish the Horndeksi, GLPV, XG3 or more exotic theories from each other. This will be the subject of future publications.


\acknowledgments
T.F.  acknowledges the supported by JSPS Postdoctoral fellowship for Research Abroad (Grant No. 27-154).
X.G. was supported by JSPS Grant-in-Aid for Scientific Research No. 25287054 and 26610062.
The work of J.Y.\ was supported by JSPS Grant-in-Aid for Scientific
Research No.\ 15H02082, and the JSPS Grant-in-Aid for Scientific
Research on Innovative Areas No.\ 15H05888.

\appendix

\section{Coefficients in (\ref{S2S})} \label{sec:Ci}

Various coefficients in (\ref{S2S}) are given by
\begin{eqnarray}
m_{AA} & = & \frac{1}{2}d_{0}+3d_{0}'+d_{0}''+3\left(a_{0}'+a_{0}''\right)H+3\left(\lambda_{1}-\lambda_{1}'+\lambda_{1}''\right)H^{2}\nonumber \\
&  & +3\left(4\lambda_{2}-3\lambda_{2}'+\lambda_{2}''\right)H^{3}+\left[d_{4}+\left(3a_{5}+a_{8}\right)H\right]\frac{k^{2}}{a^{2}}, \label{mAA}
\end{eqnarray}
\begin{equation}
g_{\zeta\zeta}=\frac{1}{2}\frac{\partial^{2}\Gamma_{1}}{\partial H^{2}},
\end{equation}
\begin{equation}
w_{\zeta\zeta}=2\left[\Gamma_{2}-\frac{1}{\bar{N}a}\frac{\mathrm{d}}{\mathrm{d}t}\left(a\frac{\partial\Gamma_{2}}{\partial H}\right)\right]+2\left[8d_{2}+3d_{3}+\left(24a_{3}+9a_{4}+8a_{6}+3a_{7}\right)H\right]\frac{k^{2}}{a^{2}},
\end{equation}
\begin{equation}
w_{BB}=b_{1}+b_{2}+\left(9c_{1}+5c_{2}+3c_{3}\right)H,
\end{equation}
\begin{equation}
f_{A\zeta}=\frac{\partial\bar{\mathcal{E}}_{A}}{\partial H},
\end{equation}
\begin{equation}
w_{A\zeta}=4\left[d_{1}-\left(9b_{3}+3b_{4}+3b_{5}+b_{6}\right)H^{2}+\Gamma_{2}'\right],
\end{equation}
\begin{equation}
w_{B\zeta}=4a_{1}+2a_{2}+2\left(12b_{3}+4b_{4}+5b_{5}+2b_{6}\right)H,
\end{equation}
where in the above, $\lambda_{1}$ and $\lambda_{2}$ defined in (\ref{lambda_1}) and (\ref{lambda_2}), $\Gamma_{1}$ defined in (\ref{Gamma_1}), and
\begin{equation}
\Gamma_{2}\equiv d_{1}+\left(3a_{1}+a_{2}\right)H+\left(9b_{3}+3b_{4}+3b_{5}+b_{6}\right)H^{2},
\end{equation}
$\bar{\mathcal{E}}_{A}$ defined in (\ref{bgeom_A}).


\providecommand{\href}[2]{#2}\begingroup\raggedright\endgroup

\end{document}